# Understanding all-dielectric periodically modulated coatings for normal-incidence polarization control

**LINA GRINEVICIUTE[1,2*], JULIANIJA NIKITINA[1] AND KESTUTIS STALIUNAS[2,3,4]**

[1]*Center for Physical Sciences and Technology (FTMC), Savanoriu Ave. 231, LT-02300 Vilnius, Lithuania*

[2]*Vilnius University, Faculty of Physics, Laser Research Center, Sauletekio Ave. 10, Vilnius, Lithuania*

[3]*ICREA, Passeig Lluís Companys 23, 08010, Barcelona, Spain*

[4]*UPC, Dep. de Fisica, Rambla Sant Nebridi 22, 08222, Terrassa (Barcelona), Spain*

**Corresponding author: lina.grineviciute@ftmc.lt*

**An ultracompact thin-film polarizer for normal-incidence (0° angle of incidence, AOI) applications is analytically and experimentally investigated. The device is based on Fano resonances in periodically nanostructured dielectric thin films, enabling polarization selective reflection and transmission due to polarization dependent resonance frequencies. The operating principle is analyzed both analytically and numerically, and the optimized structure is fabricated and experimentally characterized. Measurements demonstrate polarization contrast ratios of up to approximately 1:1000 at normal incidence. Laser-induced damage threshold measurements using nanosecond laser pulses further confirm the high damage resistance of the all-dielectric polarizer.**

The management of light polarization is crucial not only in high-tech laser systems but also in everyday technological applications. The everyday-use polarizers for incoherent light and zero angle of incidence are typically based on anisotropic metallic or polymeric structures [1]. The metallic or absorbing microwires of a fixed orientation absorb/reflect the electromagnetic radiation with the polarization directed predominantly along the wires, whereas the orthogonally polarized light is transmitted unaffected [2,3]. However, such technology cannot be used in high-power laser systems since the metallic wire-based polarizers do not stand high powers of laser radiation. Polarizers used for high-power applications are based on multilayer interference coatings, which allow polarization selectivity at high angle of incidence, such as Brewster angle [4]. Providing high contrasts and high resistivity to optical radiation, they, however, lack the compactness, as they work at non-zero angle incidence (at a Brewster angle, following the definition), and thus can hardly be integrated into the resonators of microlasers, such as microchip lasers, or semiconductor VECSELs.

Another kind of birefringence control, is based on nanostructured or sculptured thin films made by glancing angle deposition [5–8]. The application for a compact linearly polarized ceramic laser was also demonstrated by using such nanostructured coatings [9,10]. Recently sculptured silica thin films were used as high-power coatings for polarization and phase control [11]. Such coatings demonstrate very high resistivity to laser radiation but on the other hand, such polarizers provide relatively low polarization contrast, typically, of order 1:10, whereas sophisticated laser systems frequently require 1:1000 contrast. Moreover, such porous structures in principle are not stable then surrounding environment is changing. Overall, micro-laser technologies still require compact, normal-incidence polarizers that provide high polarization contrast and high laser damage resistance.

In this work, we explore and demonstrate ultra-compact, zero-angle polarizers, based on a different physical principle than [5-10], i.e. on Fano resonances in nano-modulated thin films. Fano resonances are widespread in nature and technology. After their first observation around 90 years ago [12] and their universal mathematical description 60 years ago [13], Fano resonances have found applications in different physical systems. Especially, the growth of interest in Fano resonances was boosted by rapidly developing micro- and nano-photonic technologies during the last decade [14–17]. In general, resonant nanostructure arrays have found various applications, including plasmonic and photonic crystals that control the flow of light [18] and dielectric guided-mode resonant or metasurface structures enabling enhanced nonlinear optical processes [19]. Recently, Fano resonances in dielectric thin films were proposed for efficient spatial (angular) filtering of near-zero angle incidence radiation [20,21]. The Fano resonances (in this context also called guided wave resonances), have been also proposed as polarizers at near-zero-incidence angle [22,23]. The polarization control has been investigated in similar structures, such as silicon-doped cylindrical arrays [24], dual-GaAs nanogratings [25], Si/InAs/Ag metamaterials [26], I-shaped multi-purpose splitters [27], nanodisk arrays [28], and others. See also a recent review on the topic [29]. We note that most of the theoretical studies rely on the numerical simulations of specific designs of the polarizers. The main focus of our article is to understand the core physical mechanisms behind the polarizers, to uncover the main ingredients, and, based on that understanding, to provide the analytical estimations. These main physical ingredients are two: 1) the Fano resonances are shifted in frequency one from another for different polarizations; 2) the resonances have of very different widths for different polarizations. Moreover, the article reports on the fabrication of such polarizers, experimental measurement and characterization of resistivity to laser radiation (LIDT).

The design of our structure of polarizers is a waveguiding structure, consisting of a high-index dielectric coating layer on a periodically patterned surface. A suitable substrate to provide submicronic periodicity is a fused silica (FS) as used in our design. In Fig. 1, we present the simulated and experimental transmission spectra of such a bare FS grating structure, i.e., a periodic surface pattern without any coating. The spectra are shown for p- and s-polarized light at normal incidence. Both simulation and experiment demonstrate that the transmission for the two orthogonal polarizations remains relatively similar across the entire spectral range. This indicates that the uncoated grating alone does not provide a strong polarization effect and therefore cannot serve as an efficient polarizing element at normal incidence.

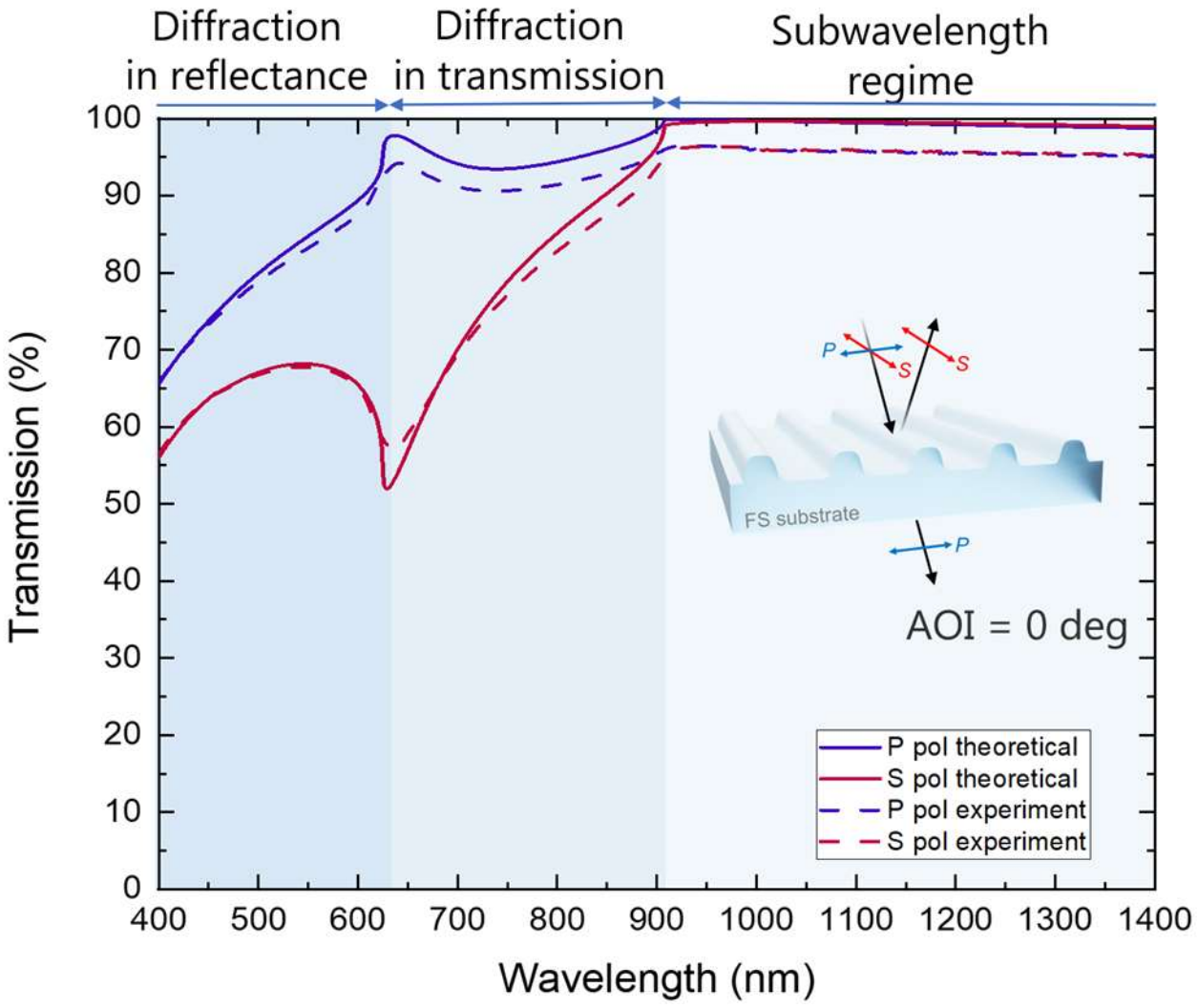

Fig. 1. Simulated and experimental transmission spectra of a periodically structured surface under normal angle of incidence (AOI = 0°) for p- and s-polarized light. The structure parameters are: $d_x$ = 625 nm,
$h_0$ = 200 nm, and $n_{sub}$ = 1.47. Different spectral regions correspond to diffraction in reflection, diffraction in transmission, and the subwavelength regime.

The main idea of a high-contrast polarizing element is depicted in Fig. 2 (a), where the interaction of incident radiation with planar waveguiding modes of the thin film is shown. The idea is based on the observation that the Fano resonances in such thin films occur at slightly different frequencies for different polarizations of the incident radiation, i.e., whether the polarization is directed along-, or perpendicular to the linear surface-modulation structure. It is known that the propagation constants of the TM and TE polarized planar modes (also called s- and p-polarized waves, respectively) in thin films are different. Hence, the excitation frequencies of these resonant modes are different as well, and eventually the reflection/transmission bands for both polarizations are displaced in frequency. We note that by tuning over the Fano resonance the reflection changes from exactly 0 till 100 % and back to zero. This provides a potential possibility to build the zero angle of incidence polarizers of extremely high contrast. Moreover, the spectral width of the resonances can be different for different perpendicular polarizations, which provides one more degree of freedom to manage the transmission/reflection. All this knowledge enabled to engineer, and to realize the normal incidence angle polarizers of high contrast, which is the main message of this letter.

First, we provide analytical estimations of these main characteristics of Fano resonances: their resonant frequencies and their spectral widths for s- and p-polarizations. Next, a detailed numerical study of light propagation through the nanostructured thin film was performed to validate the analytical predictions and identify the optimal architecture. Based on the simulation results, the structures were fabricated. Finally, transmission/reflection measurements of the fabricated structures were carried out, demonstrating high polarization contrast, and the laser-induced damage threshold (LIDT) was characterized.

**Analytic estimations**.

Here we estimate two most important characteristics of the resonances for both polarizations: their frequencies and the width of the resonance lines.

The resonances indicate the Fano resonance of the incident plane wave with the planar waveguiding mode of the thin film. If at least one surface of the thin film is periodically modulated with a period of order within a wavelength of the incident radiation, the Fabry Perot (FP) radiation can be coupled to the planar modes of the film waveguide. The planar structure (unperturbed by modulation) supports the waveguiding modes with the propagation wavenumber $k_m$ for a given frequency $\omega_0 = ck_0$ (here $k_0 = 2\pi/\lambda$ is the wavenumber of incident light in vacuum). The resonant coupling to the right/left propagating modes occurs for such incidence angles $\theta$, that $k_0 sin(\theta) \pm q_x = \pm k_m$. Here modulation wavenumber is $q_x = 2\pi/d_x$ and the period of the film modulation is $d_x$. The propagation wavenumbers of the planar modes $k_m = \sqrt{k_0^2 n^2 - q^2}$, depend on the form-factor of the planar mode $q$, and the refraction index of the material of the film $n$. $k_m$ does not have explicit analytic expression, rather obeys the transcendental relations:

$$tan(qd_z/2) = \alpha\sqrt{\frac{k_0^2}{q^2}(n^2-1) - 1} \quad (1)$$

(1) is valid for the odd modes; for the even modes $tan(qd_z/2)$ is to be substituted by $-cot(qd_z/2)$. $d_z$ is the film thickness. The coefficient $\alpha$ in (1) depends on the polarization: $\alpha = 1$ for TE mode (or s-mode), with the vector of electric field directed perpendicularly to the plane of planar waveguide (parallel to the grating lines), and $\alpha = n^2$ for the TM mode (or p-mode), with the electric vector in the waveguiding plane (perpendicular to the grating lines).

In the limit of deep (but not for infinitely deep) potential well, when the refraction index of the material of the film is high, the transverse wavenumber $q$ has an explicit approximate expression, obtained from (1) by series expansion: $tan(x) \approx 1/(m\pi/2 - x) + \cdots$, for $x \to m\pi/2$,

$$q \approx \frac{\pi m}{d_z}\left(1 - \frac{2}{\alpha d_z m k_0\sqrt{(n^2-1)}}\right) \quad (2)$$

Which is now valid both for the odd: *m=1,3,5,…* and even: *m=2,4,6,…* modes. The derivation of (2) assumes the condition $|\pi m/d_z - q| \ll k_0$.

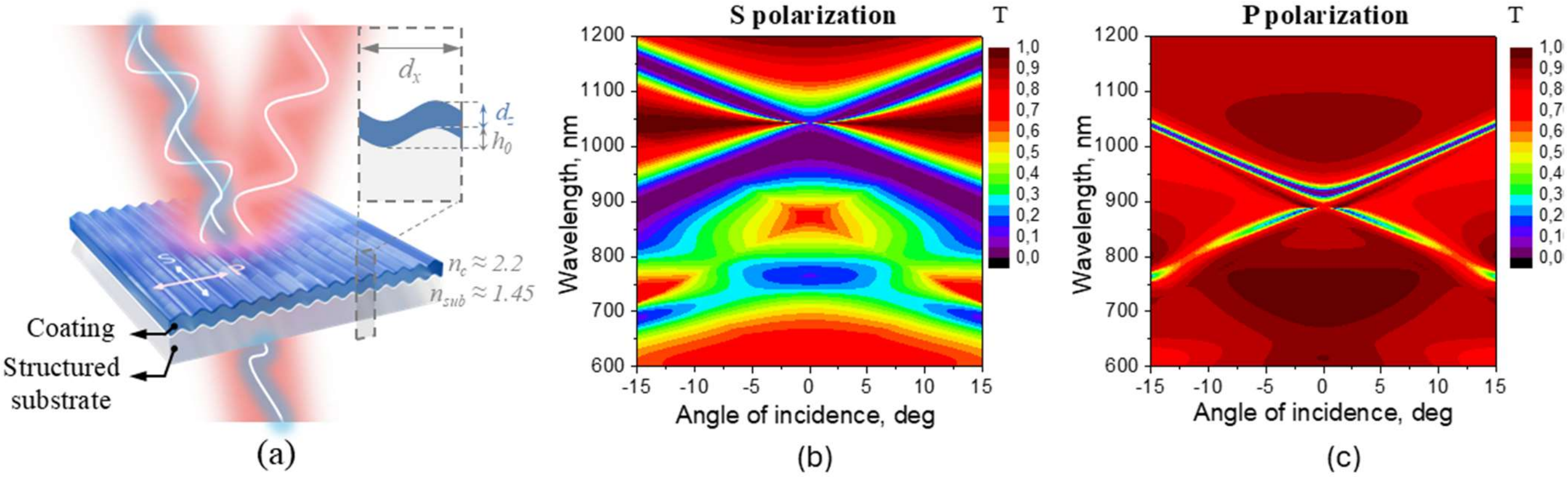

Fig. 2. a) Schematic representation of the geometry for the resonant coupling between the normal incident waves and planar waveguiding modes: the incident radiation at normal angle $\theta$ couples with left/right propagating waveguiding modes with $+k_m$ or $-k_m$. (b, c) Transmission maps in the space of the angle of incidence and the wavelength $(\theta, \lambda)$ for s- and p-polarization of the incident wave, respectively. The parameters for the calculated maps are: $d_x = 600\ nm$ , $d_z = 150\ nm$, $h_0 = 200\ nm$, and $n = 2.37$.

The transverse wavenumbers are different for different polarizations TE and TM of the planar mode, due to the factor $\alpha$. Subsequently, the propagation wavevectors $k_m = \sqrt{k_0^2 n^2 - q^2}$ are also different, which results in difference between TE and TM mode resonance frequencies. In virtue of: $(k_0 sin(\theta) \pm q_x)^2 = k_0^2 n^2 - q^2$, the resonant "cross" pattern is obtained in the parameters space of the incidence angle and wavenumber $(\theta, \lambda)$ for each planar mode and for each polarization, where the sign $\pm$ attributes the left/right inclined resonance lines, or, equivalently, the left/right propagating planar mode.

Using the above relations, one can obtain an implicit relation for the position of the resonant "cross" pattern in the parameters space of the incidence angle and wavenumber $(\theta, \lambda)$. This implicit relation also can be simplified using the approximations of strong localization. The derivations, see Supplement material, leads to:

$$\lambda_{res} \approx \frac{n d_x}{\sqrt{1 + (m d_x/(2d_z))^2}} - \frac{n^2 d_x^4}{4\alpha\pi d_z^3 \sqrt{n^2-1}} \tag{3}$$

The first part is independent on polarization, and the second is polarization sensitive due to the factor $\alpha$.

The system of the resonances is shown for both TE and TM polarizations in Fig. 2, as calculated by a commercial solver of Maxwell-Bloch equations MC Grating (see also [30,31]). In accordance with the above approximate analytical relation (3), the resonant crosses for different polarizations are displaced vertically. The amount of displacement depends on the parameters of the structure and can be varied in a large range. For the efficient zero-incidence angle (AOI=0°) polarizers, it is important that the displacement is equal to or larger than the width of the resonances. As estimated in the supplementary material, the displacement is equal to:

$$\Delta\lambda_{res} \approx \frac{d_x^4 \sqrt{n^2-1}}{4\pi d_z^3} \tag{4}$$

Another important parameter is the width of the resonance lines $\Delta\lambda_{width}$, which is proportional to the square of the coupling coefficient between the incident radiation and the waveguiding mode: $\Delta\lambda_{width}/\lambda \sim \gamma^2$. The coupling coefficient also can be estimated analytically [12] as: $\gamma \sim (n-1)h_0/\lambda$, for $h_0/\lambda \ll 1$, where $h_0$ is the amplitude of the interface modulation. The proportionality coefficient is, however, very different for both polarizations. We estimated the coupling coefficient analytically for the harmonically modulated interface between air and the material with refraction index $n$. The estimations maximally simplify around the incidence wavelength $\lambda_0 = n d_z$, see the Supplement for derivations. The resonance halfwidth (proportional to the square of the coupling coefficient) for p-polarized and s-polarized light reads:

$$\Delta\lambda_{width,s} \approx \frac{n-1}{n+1} \frac{h_0^2}{d_x} \tag{5.a}$$

$$\Delta\lambda_{width,p} \approx 2n^4 \left(\frac{n-1}{n+1}\right)^2 \frac{h_0^2}{d_x^2} (\lambda - d_z/n) \tag{5.b}$$

These estimations hold for $h_0/\lambda \ll 1$ close to the $\lambda_0 = d_z/n$, point (from the left).

The main conclusion from (5) is that the resonance widths can be very different for s- and p-polarizations, especially around $\lambda_0 = d_z/n$ point. At that point, in particular, the linewidth for p-polarization $\Delta\lambda_{width,p} \to 0$, whereas $\Delta\lambda_{width,s}$ remains finite. It also follows from (3,4) that the values of displacement and the width of resonance can be manipulated independently by proper choice of $h_0$ and $d_z$ (also of $n$ and $m$), which indicates the principal possibility to displace the resonances over more than their width. The above analysis is of an indicative character, as it is valid in the limit $h_0/\lambda \ll 1$, whereas the case of our experimental/numerical study is $h_0/\lambda \sim 1$. In this case the widths of the resonances is to be found from the numerical study.

The maps shown in the Fig. 2. are in a complete accordance with the above analytical estimations, and are at the root of the idea of the zero angle polarizers. The transmission gaps for TM and TE polarizations can be shifted one with respect to each another by the selection of proper parameters of the design (refraction index, film thickness). Therefore, non-overlapping reflection gaps can be obtained. As the transmission/reflection through the thin film within each gap ranges precisely from 0 to 100 %, high contrast polarizers can be in principle designed by matching the 0 % transmission point for one polarization with the 100 % transmission point for the other polarization, and vice versa.

**Numerical Simulations.**

We performed numerical simulations in order to test the above described analytical estimations, as well as to optimize the structures. For high contrast polarizers we searched for a case where the maximum (100 %) reflection for one polarization coincides in frequency with 0 % reflection for the other polarization at normal incidence.

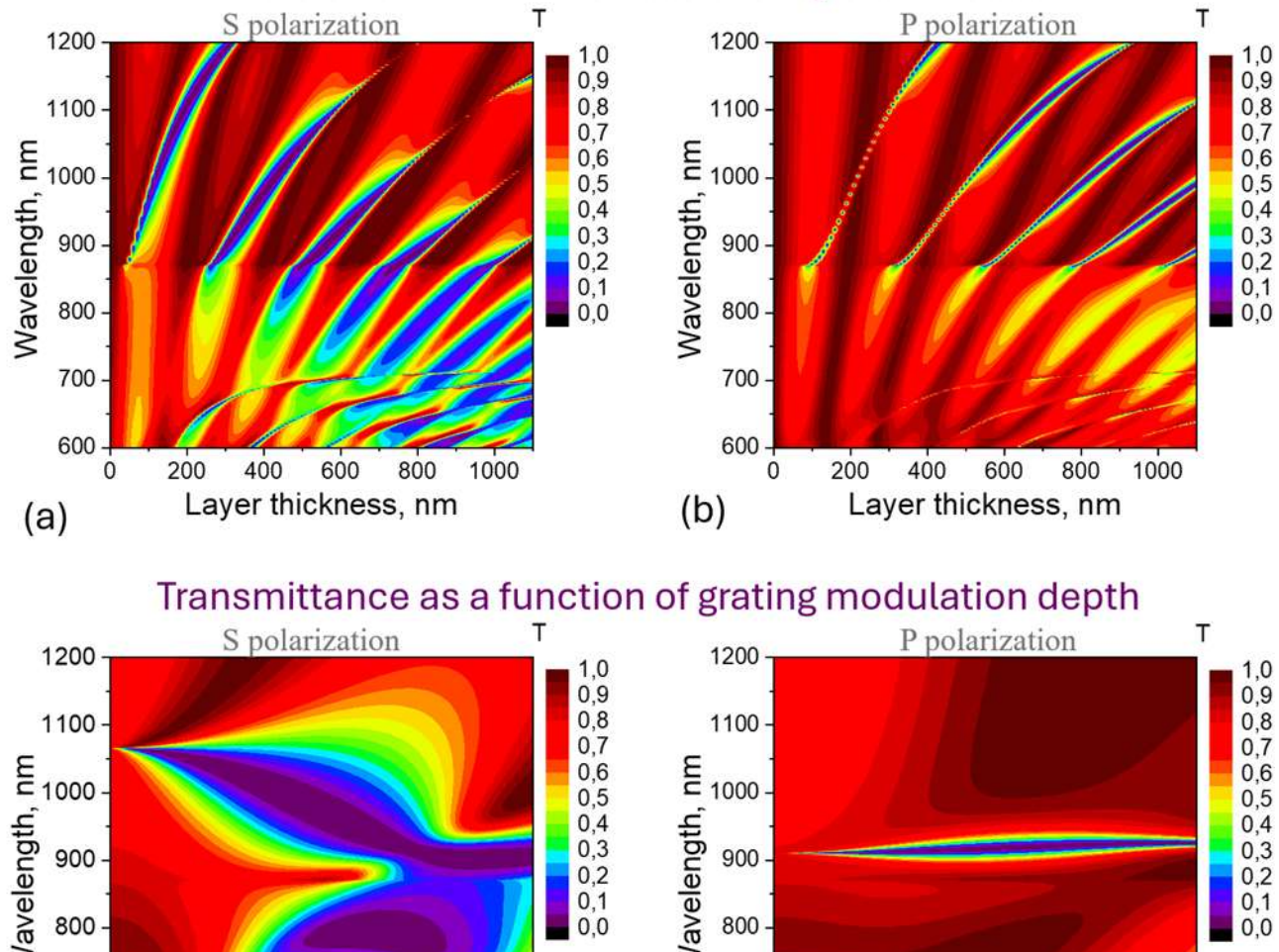

Fig. 3. Parameter analysis for fixed grating periodicity and refractive indexes ($d_x = 600\ nm$, $n_c = 2.37$ and $n_{sub} = 1.45$). (a, b) Normal-incidence transmission spectra as a function of film thickness for s- and p-polarizations, respectively ($d_x = 600\ nm$, $h_0 = 200\ nm$, $n = 2.37$). (c, d) Transmittance and resonance linewidth as a function of modulation depth ($d_z = 150\ nm$).

Dependence of the polarizing properties of nanostructured dielectric films on their geometrical parameters is shown in Fig. 3. It qualitatively verifies the predictions that the resonance positions and shape are different for perpendicular polarizations and depend on various parameters. Fig. 3 (a) and (b) show the calculated normal-incidence transmission spectra as a function of film thickness for s- and p-polarizations, respectively. Resonances for s-polarization are at larger wavelength than p-polarization, in accordance to (3,4). Fig. 3 (c) and (d) present the transmittance and corresponding resonance linewidth as functions of the grating modulation depth (other parameters are provided in the caption). It also verifies that the resonance lines for *s* are significantly broader than for the p-polarization, in accordance with (5). The results demonstrate that the modulation depth and the film thickness have a strong influence on both the position and the linewidth of the resonance, providing a way to tailor polarization contrast in such structures. Based on this analysis, we identified the optimal parameters for maximizing polarization contrast, with a target value of 1:1000.

**Fabrication and characterization**.

The polarizer was fabricated on fused silica ($n_{Sub}$ = 1.45) harmonic gratings, which served as the substrate and defined the surface corrugation for the deposited layer. The grating period was 625 nm with a modulation depth of approximately 220 nm. A single-layer thin film of titanium dioxide ($TiO_2$, $n_H$ = 2.328), a high-index dielectric material enabling the excitation of Fano resonances, was deposited using ion-beam sputtering technology with optimized process parameters. The optical constants of the materials were determined from supplementary experiments on flat-substrate test samples. The same deposition method was then applied to the modulated grating substrates to form the nanostructured thin films (see [32] for further fabrication details).

Transmittance maps for the fabricated sample were measured with spectrophotometer Photon RT using linearly polarized illumination. Two perpendicular polarizations were used: *s*- and p-, where s-polarization is parallel to the grating lines on the sample. The angle between the normal of the grating surface and the detector varied from 0° to +/- 10° in steps of 1°.

Fig. 4 summarizes the experimental realization of a metal–oxide polarizer for normal-incidence (0° AOI) applications. Fig. 4 (a) shows a photograph of the fabricated sample with indicated polarization directions, together with an SEM cross-section confirming the quality and periodicity of the grating structure. Graph (b) shows a direct comparison of the transmittance at the operational wavelength of 1064 nm, highlighting the strong polarization contrast achieved by the device. In this graph, the simulation results are compared with the experimental measurements. Simulated transmittance values at 1064 nm are 0.0022 % and 98.12 % for s- and p-polarizations, respectively. The measured transmittance values are 96.22 % and 0.01 % for p- and s-polarizations, respectively. These experimental results correspond to an averaged polarization contrast of approximately 1:1000, with a more precise extinction ratio of 1:962. The discrepancy in transmittance for p-polarization is attributed to back reflections from the rear surface of the glass sample, which are approximately 4 %. (c) and (d) display the measured transmittance maps for s- and p-polarizations, respectively, demonstrating distinct spectral responses depending on polarization.

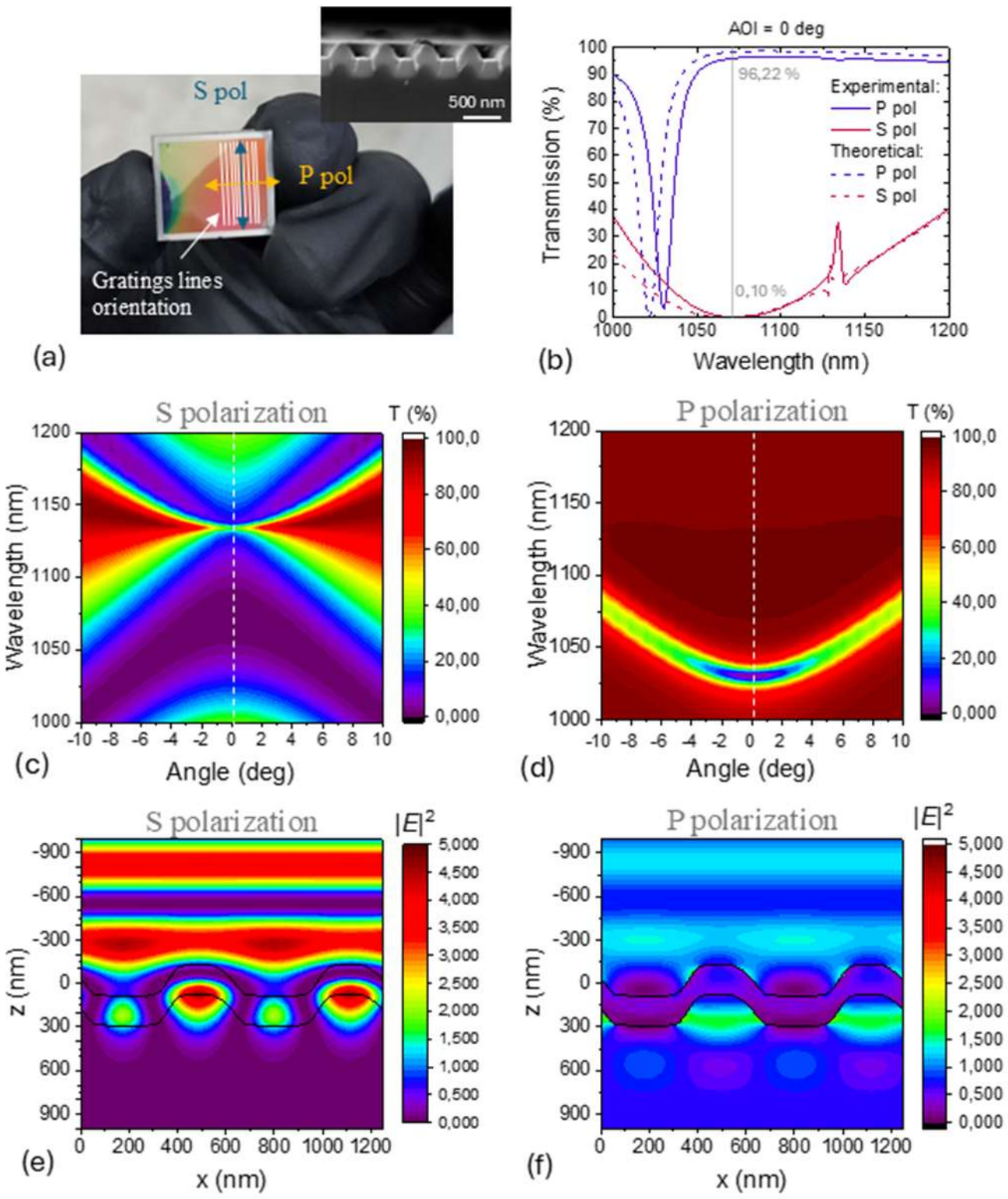

Fig. 4 Experimental realization. (a) Photograph of the sample showing polarization directions and an SEM cross-section of the fabricated structure; (b) Transmittance at 0° angle of incidence for s- and p-polarizations with precise experimental values indicated. (c,d) Measured transmittance maps for s- and p-polarizations, respectively, with marked 0° AOI cut. (e, f) Simulated electric field distribution within single-layer structure for perpendicular polarizations for 1064 nm wavelength.

Another important aspect is the distribution of the electric field within the structure. Strong field localization can lead to significant intensity enhancement, which increases the risk of structural damage and lowers the laser-induced damage threshold (LIDT). In general, the intensity of a Fano mode is inversely proportional to its resonance linewidth: the narrower the resonance, the stronger the field enhancement. To mitigate this issue, we designed the structure to operate at the resonances of s-polarization, where the resonances are broader and the field enhancement remains moderate. At the same time, this spectral region corresponds to off-resonance conditions for p-polarization, where the field enhancement is moderate. This design strategy contributes to improved laser damage resistance, which is supported by our LIDT measurements showing values of 0,7 J/cm² for s-polarization and 3,1 J/cm² for p-polarization. Measurements were achieved with 1064 nm wavelength laser (Ltd. Ekspla) in the nanosecond regime, 1-on-1 measurement protocol (pulse duration: 3 ns, spot diameter: 200 $\mu$m). These LIDT values are significantly higher than those reported for commercial 0° AOI polymeric or wire-grid polarizers, which can be attributed to the use of fully inorganic, non-absorbing materials in the polarizer structure [33,34]. The field distribution for the structure selected for fabrication is shown in Fig. 4.

## Conclusions.

The main conclusion is that an ultracompact thin-film polarizer has been numerically and analytically investigated, designed, and experimentally demonstrated. The polarizer exhibits theoretically unlimited contrast for monochromatic plane waves at normal incidence, while experimentally achieving values as high as approximately 1:1000. Moreover, laser-induced damage threshold (LIDT) measurements were performed using nanosecond laser pulses, demonstrating high values for a normal-incidence (0° AOI) polarizer.

The demonstrated single-layer dielectric polarizers already provide high contrast at normal angle of incidence; however, their performance can be further improved by extending the concept to multilayer coating designs. While the fundamental polarization mechanism remains the same, adding additional layers introduces new degrees of freedom, such as layer thickness and refractive index modulation that can be exploited to optimize angular and spectral tolerance.

Moreover, multilayer designs can enhance the structural robustness of the polarizers and increase their resistance to laser-induced damage. Due to their ultracompact design, high contrast, and potential scalability, these structures represent a highly promising class of optical elements for integration into high-power microlasers systems. Very recent attempts to integrate such polarizers into resonator of microchip laser were readily reported [35].

## AUTHORS' CONTRIBUTIONS

To L.G. belongs the idea, simulations and supervision of experimental research. J.N. fabricated the structure and performed measurements. K.S. performed an analytical study. All authors contributed to the interpretation of the results and preparation of the article.

## ACKNOWLEDGEMENTS

This project has received funding from the Research Council of Lithuania (LMTLT), agreement No. S-PD-24-94. K.S. was supported by Spanish Ministry of Science, Innovation and Universities (MICINN) under the project PID2022-138321NB-C21.

## DATA AVAILABILITY

The data that support the findings of this study are available from the corresponding author upon reasonable request.

## REFERENCES

[1] Yeh P. A new optical model for wire grid polarizers. Opt Commun 1978;26:289–92. https://doi.org/10.1016/0030-4018(78)90203-1.

[2] Wang C, Chao Y, Liang J, et al. Large-Area Wire Grid Polarizer with High Transverse Magnetic Wave Transmittance and Extinction Ratio for Infrared Imaging System. Adv Photonics Res 2023;4:2200218. https://doi.org/10.1002/ADPR.202200218.

[3] Zhao Z, Zhao Z, Ma T, et al. Numerical demonstration of low-reflective wire grid polarizers with a patterned $Fe_2O_3$ absorptive layer. Applied Optics, Vol 61, Issue 32, Pp 9708-9715 2022;61:9708–15. https://doi.org/10.1364/AO.472299.

[4] Zhu M, Yi K, Fan Z, et al. Theoretical and experimental research on spectral performance and laser induced damage of Brewster's thin film polarizers. Appl Surf Sci 2011;257:6884–8. https://doi.org/10.1016/J.APSUSC.2011.03.023.

[5] Grineviciute L, Moein T, Han M, et al. Optical anisotropy of glancing angle deposited thin films on nano-patterned substrates. Opt Mater Express 2022;12:1281. https://doi.org/10.1364/ome.451669.

[6] Oliver JB, Kessler TJ, Smith C, et al. Electron-beam–deposited distributed polarization rotator for high-power laser applications. Opt Express 2014;22:23883. https://doi.org/10.1364/oe.22.023883.

[7] Oliver JB, Smith C, Spaulding J, et al. Glancing-angle–deposited magnesium oxide films for high-fluence applications. Opt Mater Express 2016;6:2291. https://doi.org/10.1364/ome.6.002291.

[8] Mireles M, Hoffman BN, MacNally S, et al. Direct-write laser-assisted patterning of form birefringence in wave plates fabricated by glancing-angle deposition. Optica 2023;10:657. https://doi.org/10.1364/optica.487263.

[9] Doucet A, Beydaghyan G, Ashrit P V., et al. Compact linearly polarized ceramic laser made with anisotropic nanostructured thin films. Applied Optics, Vol 54, Issue 28, Pp 8326-8331 2015;54:8326–31. https://doi.org/10.1364/AO.54.008326.

[10] Lakhtakia A, Hodgkinson I, Messier R, et al. Linear and Circular Polarization Filters Using Sculptured Thin Films. Optics and Photonics News, Vol 10, Issue 12, Pp 30-31 1999;10:30–1. https://doi.org/10.1364/OPN.10.12.000030.

[11] Grineviciute L, Ramalis L, Buzelis R, et al. Highly resistant all-silica polarizing coatings for normal incidence applications. Opt Lett 2021;46:916. https://doi.org/10.1364/ol.414392.

[12] Fano U. Sullo spettro di assorbimento dei gas nobili presso il limite dello spettro d'arco. Il Nuovo Cimento (1924-1942) 1935 12:3 2008;12:154–61. https://doi.org/10.1007/BF02958288.

[13] Fano U. Effects of Configuration Interaction on Intensities and Phase Shifts. Physical Review

1961;124:1866. https://doi.org/10.1103/PhysRev.124.1866.

[14] Limonov MF, Rybin M V., Poddubny AN, et al. Fano resonances in photonics. Nature Photonics 2017 11:9 2017;11:543–54. https://doi.org/10.1038/nphoton.2017.142.

[15] Stern L, Grajower M, Levy U. Fano resonances and all-optical switching in a resonantly coupled plasmonic-atomic system. Nat Commun 2014;5:1–9. https://doi.org/10.1038/NCOMMS5865;TECHMETA.

[16] Campione S, De Ceglia D, Guclu C, et al. Fano collective resonance as complex mode in a two-dimensional planar metasurface of plasmonic nanoparticles. Appl Phys Lett 2014;105. https://doi.org/10.1063/1.4901183.

[17] Yang Y, Wang W, Boulesbaa A, et al. Nonlinear Fano-Resonant Dielectric Metasurfaces. Nano Lett 2015;15:7388–93. https://doi.org/10.1021/ACS.NANOLETT.5B02802.

[18] Collin S. Nanostructure arrays in free-space: Optical properties and applications. Reports on Progress in Physics 2014;77. https://doi.org/10.1088/0034-4885/77/12/126402.

[19] Raghunathan V, Deka J, Menon S, et al. Nonlinear optics in dielectric guided-mode resonant structures and resonant metasurfaces. Micromachines (Basel) 2020;11. https://doi.org/10.3390/MI11040449.

[20] Grineviciute L, Nikitina J, Babayigit C, et al. Fano-like resonances in nanostructured thin films for spatial filtering. Appl Phys Lett 2021;118. https://doi.org/10.1063/5.0044032.

[21] Lukosiunas I, Grineviciute L, Nikitina J, et al. Extremely narrow sharply peaked resonances at the edge of the continuum. Phys Rev A (Coll Park) 2023;107. https://doi.org/10.1103/PhysRevA.107.L061501.

[22] Lee KJ, Giese J, Ajayi L, et al. Resonant grating polarizers made with silicon nitride, titanium dioxide, and silicon: Design, fabrication, and characterization. Opt Express 2014;22:9271. https://doi.org/10.1364/oe.22.009271.

[23] Hemmati H, Bootpakdeetam P, Magnusson R. Metamaterial polarizer providing principally unlimited extinction. Opt Lett 2019;44:5630. https://doi.org/10.1364/ol.44.005630.

[24] Frang Jimin, Wang Bo. Silicon-doped cylindrical arrays for ultra-broadband terahertz absorber with polarization independence. Phys Scr 2021;96. https://doi.org/10.1088/1402-4896/ac0076.

[25] Shi X, Wang B. Dual-polarization strong nonreciprocal radiation by the 2D GaAs nanograting. Appl Phys Lett 2024;125:62202. https://doi.org/10.1063/5.0225127/3307378.

[26] Li J, Wang B, Wu J. Si/InAs/Ag metamaterial for strong nonreciprocal thermal emitter with dual polarization under a 0.9 T magnetic field. Appl Mater Today 2024;39:102345. https://doi.org/10.1016/J.APMT.2024.102345.

[27] Xiong Z, Wang B. I-shaped stack configuration for multi-purpose splitter. Opt Laser Technol 2024;168:109959. https://doi.org/10.1016/J.OPTLASTEC.2023.109959.

[28] Huang Z, Wang B. High-efficiency multi-port beam control device based on periodic nanodisk arrays. Opt Laser Technol 2022;152:108102. https://doi.org/10.1016/J.OPTLASTEC.2022.108102.

[29] Xie Y, Chen C, Wang J, et al. Guided Mode Resonant Gratings for Broadband Reflection. Laser Photon Rev 2025. https://doi.org/10.1002/lpor.202501956.

[30] Chandezon J, Dupuis MT, Cornet G, et al. Multicoated gratings: a differential formalism applicable in the entire optical region. J Opt Soc Am 1982;72.

[31] Modal and C Methods Grating Software n.d. https://mcgrating.com/ (accessed September 19, 2025).

[32] Grineviciute L, Melnikas S, Nikitina J, et al. Optical Coating Deposition on Submicron-Patterned Surfaces. Coatings 2025;15. https://doi.org/10.3390/coatings15040372.

[33] Polarizers – absorptive, polarizing beam splitters, birefringence, calcite, Glan–Taylor prism, Wollaston prism, thin-film polarizers n.d. https://www.rp-photonics.com/polarizers.html?utm_source=chatgpt.com (accessed January 18, 2026).

[34] Thorlabs Polymer Linear Polarizer - LPNIRE050-B | MEETOPTICS n.d. https://www.meetoptics.com/polarization-optics/linear/polymer-linear-polarizer/s/thorlabs/p/LPNIRE050-B (accessed January 18, 2026).

[35] Plukys M, Grineviciute L, Nikitina J, et al. Enhancement of brightness in microchip laser with angular filtering mirrors. Opt Laser Technol 2025;181. https://doi.org/10.1016/j.optlastec.2024.111904.